\documentclass[trans]{IEEEtran}
\usepackage{amsfonts}
\usepackage{amssymb}
\usepackage{amsmath}
\usepackage{amsthm}
\usepackage{cases}
\usepackage{nccbbb}
\usepackage{epsfig}
\usepackage{booktabs}
\usepackage{graphicx,subfigure}
\usepackage[ruled, linesnumbered]{algorithm2e}
\usepackage{url}
\usepackage{multirow}
\usepackage{multicol}
\usepackage[font=footnotesize,labelfont=bf]{caption}
\usepackage{cite}
\usepackage{lineno}
\usepackage{rotating}
\graphicspath{{./Figure/}}

\begin{document}
	
\title{Toward Open-World Electroencephalogram Decoding Via Deep Learning: A Comprehensive Survey}

\author{Xun Chen, Chang Li, Aiping Liu, Martin J. McKeown, Ruobing Qian, Z. Jane Wang}

\maketitle

Electroencephalogram (EEG) decoding aims to identify the perceptual, semantic, and cognitive content of neural processing based on non-invasively measured brain activity. Traditional EEG decoding methods have achieved moderate success when applied to data acquired in static, well-controlled lab environments. However, an open-world environment is a more realistic setting, where situations affecting EEG recordings can emerge unexpectedly, significantly weakening the robustness of existing methods. In recent years, deep learning (DL) has emerged as a potential solution for such problems due to its superior capacity in feature extraction. It overcomes the limitations of defining `handcrafted' features or features extracted using shallow architectures, but typically requires large amounts of costly, expertly-labelled data -- something not always obtainable. Combining DL with domain-specific knowledge may allow for development of robust approaches to decode brain activity even with small-sample data. Although various DL methods have been proposed to tackle some of the challenges in EEG decoding, a systematic tutorial overview, particularly for open-world applications, is currently lacking. This article therefore provides a comprehensive survey of DL methods for open-world EEG decoding, and identifies promising research directions to inspire future studies for EEG decoding in real-world applications. 

\section{Introduction}
Identifying and predicting mental processes from observed patterns of neural activities have long been explored in cognitive neuroscience and brain computer interfaces \cite{hassan2018electroencephalography}. Noninvasive techniques, such as  electroencephelography (EEG), magnetoencephelography (MEG), near infrared spectroscopy (NIRS), and functional magnetic resonance imaging (fMRI) provide accessible ways to broadly examine brain activity without surgical intervention. In particular, the EEG is popular for brain decoding in practical applications, as it is relatively inexpensive and has advantages of safety, high temporal resolution, wide accessibility, and potential portability \cite{hubbard2019eeg, tao2020eeg}. EEG signals can be obtained by placing electrodes on the surface of the scalp, providing measurements of post-synaptic potentials -- an indirect measure of neuronal activity \cite{chen2016joint}. This allows for, after digitization and suitable analyses, decoding of brain states and communication between our brain and the outside world \cite{schirrmeister2017deep}. 

EEG decoding methods have made great progress in recent decades. However, most existing EEG decoding methods were designed using data collected in  static or well-controlled lab environments that utilized rigorous experimental protocols and strict laboratory conditions, which are unrealistic in an open-world environment. Open-world EEG decoding refers to identifying the perceptual, semantic, and cognitive content of measured brain activity in an open-world environment \cite{zhang2015eeg}. With the emergence of new open-world applications in various fields, such as in entertainment, industry, and medicine, there is an urgent need to develop efficient EEG decoding methods for real-world scenarios. Additionally, such open-world applications typically adopt few-electrode portable, wearable, and wireless systems to take advantage of technological advances in hardware \cite{minguillon2017trends}. These recordings from complex environments tend to be heavily contaminated with artifact, making brain decoding even more challenging.
 
\begin{table*} \renewcommand{\arraystretch}{1.5}
	\caption{Challenges and solutions for three typical steps of open-world EEG decoding via DL. }
	\label{table1}
	\centering
	\begin{tabular}{*{3}{|c}|}
		\hline
		& Challenges & Solutions  \\ 
		\hline
		\multirow{4}{*}{Preprocessing} & \multirow{4}{*}{How to remove various EEG artifacts automatically with good generalization ability} & CNN   \\
		\cline{3-3}
		&  & Autoencoder  \\
		\cline{3-3}
		&  & LSTM  \\
		\cline{3-3}
		&  & GAN  \\
		\hline
		\multirow{4}{*}{Feature extraction} & How to solve the distribution mismatch between the source and the target EEG data & Transfer learning   \\
		\cline{2-3}
		& \multirow{2}{*}{How to exploit complementary information of EEG from multiple tasks and modalities}  & Multi-task learning  \\
		\cline{3-3}
		&  & Multi-modal learning  \\
		\cline{2-3}
		& How to make the EEG models robust for adversarial attacks  & Adversarial training  \\
		\hline
		\multirow{5}{*}{Classification}& \multirow{2}{*}{How to design EEG models with desirable performance under small sample size} & Few-shot learning  \\
		\cline{3-3}
		&   & Semi-supervised learning   \\
		\cline{2-3}
		& How to recognize both the known and the unknown categories of EEG  & Zero-shot learning   \\
		\cline{2-3}
		& How to exploit the structure of unlabeled EEG data to provide supervision & Self-supervised learning   \\
		\cline{2-3}
		& How to train robust EEG models in the presence of noisy label & Noisy label classification   \\
		\hline
	\end{tabular}
\end{table*}

The steps involved in EEG decoding typically include preprocessing, feature extraction, and classification, and successful open-world EEG decoding requires specific considerations at each step. Even under the most stringent recording conditions, EEG signals are easily corrupted by various artifacts (\emph{e.g.}, eye blinks, muscle artifacts, cardiac interference, and electromagnetic interference) \cite{minguillon2017trends}.  EEGs in open-world environments are also contaminated by outdoor open-world artifacts caused by extensive movement (such as muscle and mechanical artifacts) and electromagnetic factors \cite{minguillon2017trends}.  Historically EEG features have been extracted from time-domain (such as mean, variance, and kurtosis), frequency-domain (such as power spectral density and fast Fourier transform), and time-frequency domains (such as discrete wavelet transform). In addition, traditionally--defined features strongly depend on human expertise in a specific domain, and manual feature extraction is time-consuming. Classification of the extracted features include such techniques such as decision tree (DT), support vector machine (SVM), and linear discriminant analysis (LDA). EEG signals are temporally non-stationary, with their statistics varying over time which can make generalization of a classifier challenging based on limited amounts of  data. 

During feature extraction, most EEG decoding methods assume that the training and test data are identically distributed. However, high inter-subject variability, electrode shifts, and physiological state changes will inevitably lead to mismatch between training and test distributions in an open-world situation  \cite{wan2021review}. Even a small mismatch in distributions can cause significant performance degradation. 

To overcome the above challenges, deep learning (DL) methods, which are an automatic end-to-end learning framework, consisting of preprocessing, feature extraction, and classification, have achieved state-of-the-art performance in the field of EEG decoding \cite{roy2019deep}, with better generalization abilities and more flexible applicability. DL approaches avoid time-consuming preprocessing and feature extraction by working on raw EEG signals directly to learn useful information, which can capture both discriminative high-level features and underlying dependencies.

Despite their successes, DL approaches have their own challenges. Supervised DL implicitly assumes that there exists a large number of labelled EEG training samples for DL to achieve good generalization performance \cite{roy2019deep}. However, EEG classification performance deteriorates as the number of available training samples diminishes, emphasizing the need for robust DL models under more typical small sample size scenarios \cite{lu2010regularized}. Transfer learning \cite{wan2021review}, multi-task learning \cite{song2019eeg} and multi-modal learning \cite{jia2021sleepprintnet} have been recently proposed to address small-sample data concerns.  DL models have usually assumed that the categories of the test EEG signals have been seen during training. However, there may be test EEG signals that do not belong to any category of the training set. Zero-shot learning \cite{duan2020zero} may be a potential solution to recognize both known and unknown categories of EEG signals. Another challenge with DL approaches is that the labels are likely not perfectly assigned in the training set \cite{porbadnigk2014brain}, resulting in overfitting of the incorrect labels and reducing ultimate classification accuracy when applied to unseen data. Unlabeled EEG data can be analyzed via self-supervised \cite{banville2020uncovering} and semi-supervised learning methods \cite{jia2014novel} to obtain important latent information, which we will expand on later.  Finally, DL models can be easily fooled with adversarial examples, which are modified normal examples with small deliberate perturbations \cite{zhang2019vulnerability}, so care must be given to prevent this possibility.

Although there are several review articles on EEG decoding, to the best of our knowledge, most of them focus on EEG decoding in static or well-controlled lab environments. In this article, we review open-world EEG decoding via DL, which has the capacity to develop robust EEG decoding for real open-world applications. Our objectives are as follows: (1) present the problems and challenges faced by open-world EEG decoding, (2) provide a taxonomy of DL solutions for open-world EEG decoding, and (3) discuss potential ways to obtain more robust DL models for open-world EEG decoding.

\section{Open-world EEG decoding via DL}
In this section, we introduce new problems and methods to deal with open-world EEG decoding via DL on each step of EEG decoding, \emph{i.e.}, preprocessing, feature extraction, and classification. The structure of open-world EEG decoding via DL is schematically illustrated in Table \ref{table1}.

\subsection{Preprocessing}
The performance of EEG decoding depends heavily on the quality of the EEG signals. Unfortunately, the recorded signals are usually contaminated by various artifacts, which can be magnified in open-world applications within complex environments \cite{minguillon2017trends}.  Therefore, it is of both theoretical and practical significance to remove complex artifacts from contaminated EEG signals in the open-world.

Filtering and regression are traditional artifact removal methods. Filtering methods assume that artifacts and EEG signals reside in distinct frequency bands \cite{chen2019removal}. However, artifacts and EEG signals usually overlap in the frequency domain, and there is a risk that portions of the EEG signals may be eliminated during this artifact removal process. Regression methods usually assume that each EEG channel can be modeled as a linear or nonlinear superposition of clean brain activity and artifact signals that can be obtained from reference channels or artifact templates. However, regression methods only work when suitable reference channels that are available (\emph{e.g.}, channels for measuring eye movement).

Another popular approach for EEG denoising during the preprocessing stage is blind source separation (BSS) \cite{jung2000removing}, which assumes that the clean EEG and artifacts are statistically independent in the time domain, so they will be isolated into different components. This allows for removal of  artifact-related components during the reconstruction process. BSS methods typically require human intervention to identify the artifact-related components, which is subjective and time-consuming. These methods generally require that the number of channels must be larger than or equal to the number of underlying sources \cite{chen2019removal}, so they are less attractive when only a few channels are available, as is typical in mobile scenarios.

Empirical mode decomposition (EMD) \cite{huang1998empirical} and the wavelet transform (WT) \cite{vazquez2012blind} are two representative methods for denoising when only a limited number of EEG channels are available. EMD decomposes an input signal into multiple empirical modes according to the intrinsic mode function (IMF). IMFs are a set of the band-limited functions that satisfy two basic conditions: (1) the number of extreme points and the number of zero crossings must either equal or differ at most by one, and (2) at every point, the mean value of the envelopes defined by local maxima and local minima should be zero. EMD's  data-driven approach capable of dealing with non-stationary stochastic processes makes it suitable for removing artifacts from contaminated EEG signals. However, EMD is time-consuming and is not suitable for online applications in the open-world. Similar to EMD, the WT first decomposes the contaminated EEG signal into different sub-bands. Then, a threshold function is used to update the coefficients related to the sub-bands that are assumed to be artifact-related. Finally, the EEG signals are reconstructed using the updated coefficients. However, selection of an incorrect threshold setting could lead to the degradation of the reconstructed EEG signals \cite{minguillon2017trends}. A traditional neural network with shallow layers can be used to replace the threshold function in the wavelet analysis, which has the advantage of approximating smooth nonlinear functions. Nevertheless, the approximation ability of a traditional, shallow layer neural network tends to be inferior to that of DL.

DL-based methods can be used to automatically filter out artifacts from contaminated EEG signals. One typical method is to learn a mapping between noisy EEG signals and their cleaned versions \cite{zhang2021eegdenoisenet}. The performance of DL-based artifact removal methods relies fundamentally on the size of the training datasets. For example, Zhang \emph{et al.} established a benchmark EEG dataset for the training and testing of DL-based artifact removal methods \cite{zhang2021eegdenoisenet}. The EEG epochs were acquired from a motor imaginary EEG dataset, with a band-pass filter between 1 and 80 Hz applied, followed by re-sampling to 256 Hz. Then, the ICLabel method \cite{pion2019iclabel} was used to attenuate the artifacts. Finally, the EEG signals were segmented into epochs of 2-$s$. The ocular artifact epochs were acquired from open-access EEG data, band-pass filtered between 0.3 and 10 Hz, followed by re-sampling to 256 Hz, then segmented into 2-$s$ epochs. The myogenic artifact epochs were acquired from a facial surface electromyography (EMG) dataset, band-pass filtered between 1 and 120 Hz, re-sampling to 512 Hz, then segmented into 2-$s$ epochs. For all the categories, the epochs were standardized by subtracting their mean and dividing by their standard deviation, and then were visually checked by an expert. Finally, 4514 clean EEG epochs, 3400 ocular artifact epochs, and 5598 muscular artifact epochs were acquired. Simulated noisy signals can be generated by linearly mixing the clean EEG epochs with EOG or EMG epochs, the SNRs for EEG epochs contaminated by ocular artifacts range from -7dB to 2dB, and the SNRs for those contaminated by myogenic artifacts range from -7dB to 4dB. In this way, the clean EEG epochs can be considered as ground truth, and the mixed epochs as contaminated EEG. This allowed the adoption of a large number of noisy EEG epochs with ground truth (clean EEG epochs) for model training and testing. Mathematically, simulated noisy EEG signals can be formulated as
\begin{equation}\label{Eq1}
\mathbf{Y} = \mathbf{X} + \lambda \cdot \mathbf{N},
\end{equation}
where $\mathbf{Y}$ denotes the contaminated EEG signal with artifacts, $\mathbf{X}$ denotes the clean EEG signal, $\mathbf{N}$ denotes the artifacts, and $\lambda$ denotes the relative contribution of the artifacts.

The goal of DL-based artifact removal is to learn an end-to-end nonlinear function $f$ to map a noisy EEG signal to approximate a clean EEG signal as follows
\begin{equation}\label{Eq2}
\hat{\mathbf{X}} = f(\mathbf{Y}, \theta),
\end{equation}
where $\hat{\mathbf{X}}$ denotes the approximated clean EEG signal, and $\theta$ denotes the parameters to be learned. The learning process can be realized by minimizing the objective function as follows
\begin{equation}\label{Eq3}
\hat{\theta} = {\text{arg}} ~ \underset{\theta} {\text{min}} ~ \frac{1}{N} \sum_{i=1}^{N} \| \mathbf{X}_i - f(\mathbf{Y}_i, \theta) \|_F^2, 
\end{equation}
where $N$ represents the number of training samples, and $\mathbf{Y}_i$ and $\mathbf{X}_i$ represent the $i^{th}$ contaminated EEG signal and clean EEG signal, respectively. After obtaining the optimized trained parameters $\hat{\theta}$, denoised EEG signals can be obtained for the EEG test dataset.

Autoencoder, one of the major branches of DL, has also been used in artifact removal \cite{ghosh2019automated}. Autoencoder consists of three layers, \emph{i.e.}, input, hidden and output layers. An autoencoder maps an input $x$ to an output $z$. It takes an input $x$ and maps it to a hidden representation $y$ through a mapping (encoder) as follows
\begin{equation}\label{Eq4}
y = s( W^T x + b),
\end{equation}
where $s$ denotes the non-linear function such as sigmoid, $W$ and $b$ denote the weight and bias vector from input to hidden layer, respectively. The hidden representation is then mapped back (decoder) again to the output $z$ of the same size as input $x$ as follows
\begin{equation}\label{Eq5}
z = s( W'^Ty + b'),
\end{equation}
where $W'$ and $b'$ denote the weight and bias vectors from hidden layer to the output layer, respectively. The training of the network can be accomplished by measuring the reconstruction error, which can be measured with the traditional mean squared error (MSE). For example, Ghosh \emph{et al.} proposed an automated eye blink artefact removal from EEG using SVM and autoencoder \cite{ghosh2019automated}. A sliding window of 0.45-$s$ is applied to the EEG data, and each window is processed as follows: (1) Identification of artifacts: The signal within the window is input to the SVM classifier, SVM classifies whether the signal is an artifact or clean EEG. If it classifies the signal as an artifact, it is input to the autoencoder for correction. On the other hand, if the signal is classified as non-artifact, then the window is slid forward. (2) Cleaning the artifacts: Signal window marked as artifact is then input to the pre-trained autoencoder. The autoencoder removes the artifact of contaminated EEG window and outputs the clean EEG signal. However, the encoder and decoder in \cite{ghosh2019automated} are the simplest case of the autoencoder, which use only one linear layer and nonlinear layer to represent both the encoder and decoder. The encoder and decoder for EEG preprocessing can have multiple layers. For example, a reversible GAN was used for the removal of BCG artifacts \cite{lin2020single}. An autoencoder is used as a subnetwork in the single shot reversible GAN. The encoder contains multiple convolutional layers, and the decoder contains multiple transposed convolutional layers (deconvolutional layers).

Long short-term memory (LSTM) models, a popular variant of recurrent neural network (RNN), which solve the vanishing and exploding gradient problem of RNN by adding extra parameters to the RNN model \cite{hochreiter1997long}, can also be used for EEG artifact reduction. An LSTM is composed of four different gates, which checks the input of the cell state and determines the influence of the cell state on the output. The four gates are termed as input gate, forget gate, output gate and block input gate. The input gate and the block input gate control the new information flow to the memory cell. Sigmoid and Tanh are used as activation functions for the input gate and block input gate, respectively. The forget gate controls which previous information should be retained into memory cell. The output gate determines what is to send as output from LSTM unit. A sigmoid activation function is used for both the forget gate and the output gate. Manjunath \emph{et al.} proposed a low complexity LSTM for detecting various artifacts in multi-channel brain EEG signals \cite{manjunath2020low}. It consists of one average pooling layer of size 64, one LSTM layer having 12 units, one dense layer having 50 neurons and one output layer for binary classification. The experimental results indicate that the LSTM based FPGA hardware outperforms the CNN based FPGA hardware by 1.88$\times$ in terms of dynamic power consumption per classification.

Although the deep network-based methods described above have achieved desirable performance, an end-to-end network typically uses the MSE between the network output and ground truth as the loss function, leading to over-smoothing and loss of detail \cite{ledig2017photo}. To overcome these problems, a generative adversarial network (GAN) can be used to remove artifacts, where a generative network is trained to map a noisy EEG signal to a clean EEG signal, and a discriminator network is trained to discriminate between real and generated EEG signals. An example where this is useful is in simultaneous EEG and functional magnetic resonance imaging (fMRI) recordings which can measure brain activity with both high temporal and spatial resolution.  A ballistocardiogram (BCG) artifact exists due to cardiac activity and blood flow inside the static magnetic field of the MRI scanner \cite{lin2020single}. However, BCG artifact removal remains challenging using DL, since it is difficult to obtain clean EEG signals and BCG-contaminated EEG signals at the same time. A paired signal-to-signal problem refers to the mapping between input data and output data using paired training data. However, it is hard to obtain BCG-contaminated EEG signals and clean EEG signals in the same state simultaneously, so BCG artifact removal can be considered an `unpaired signal-to-signal problem'. A cycle-consistent generative adversarial network (CycleGAN) \cite{zhu2017unpaired} is a  technique to solve the unpaired image-to-image translation, and it can be widely used in many applications, such as unpaired image denoising, unpaired image super-resolution, and unpaired image dehazing. Since the SNR of the EEG signal is low, the direct use of CycleGAN, which performs well on the unpaired problem, still cannot easily remove the BCG artifacts in  simultaneous EEG-fMRI. Lin \emph{et al.} proposed a novel single-shot reversible GAN for the removal of BCG artifacts \cite{lin2020single}. Being capable of bidirectional input and output, the forward model can map contaminated EEG signals to clean EEG signals, and the reverse model can achieve data conversion from clean EEG signals to contaminated EEG signals.

\subsection{Feature extraction}

\subsubsection{Transfer learning for addressing the distribution mismatch issues}
Recently,  EEG recognition methods have proven successful in many applications,  particularly when the training and test EEG data are drawn from the same distribution. However, this is not the case in many real open-world problems, where there is notable high inter-subject variability, electrode shift, and physiological state changes, all of which affect the generalization ability of models \cite{wan2021review}. The performance of a classifier trained in the source domain will almost certainly drop when tested on the target domain due to the distribution mismatch between the source and target domains, limiting its practical use.

Transfer learning aims to improve the performance of learners in the target domain by fusing knowledge from one or more related, but differently--distributed source domains. For example, it may be desired to augment small-sample EEG from one institution with large data sets collected from other institutions. Domain adaptation is a special case of transfer learning that uses labeled data in one or more source domains to improve the learning performance in a target domain. Traditional domain adaptation methods in EEG signal analysis include distribution and subspace adaptation \cite{wan2021review}. Distribution adaptation can be generally classified into marginal and conditional distribution adaptation. The objective of marginal distribution adaptation is to transfer knowledge when the marginal distributions of the source domain ($X_S$) and target domain ($X_T$) are different, \emph{i.e.}, $P(X_S) \neq P(X_T)$. The objective of conditional distribution adaptation is to transfer knowledge when the conditional distributions of the source and target domains are different, \emph{i.e.}, $P(Y_S|X_S) \neq P(Y_T|X_T)$. Measures of marginal distribution difference and conditional distribution difference include maximum mean discrepancy \cite{borgwardt2006integrating}, Kullback--Leibler divergence \cite{gupta2009classification}, and Jensen--Shannon divergence \cite{giles2019subject}. Subspace adaptation transforms data in both the source and target domains into a common latent subspace in which their distributions are similar \cite{anderson2006geometric}. Linear subspace adaptation usually utilizes linear subspace learning algorithms for domain adaptation in EEG signal analysis, such as principal component analysis and linear discriminant analysis. Manifold learning methods map the original high-dimensional EEG data in the source and target domains into a common low-dimensional manifold structure \cite{wan2021review}. Statistical feature alignment aims to map the EEG data into a subspace to align the statistical features in the source and target domains \cite{he2019transfer}, such as variance and median absolute deviation. However, traditional domain adaptation methods align the distribution in the source and target domains or learn shared common subspaces with shallow representations.

Recently, several studies have demonstrated that deep networks can learn more transferable representations \cite{wang2018deep}. The deep features eventually transition from general to specific, and the transferability sharply decreases in higher layers that are near to the output \cite{yosinski2014transferable}. The most commonly--used method in EEG signal analysis is to fine-tune a pre-trained DL model \cite{raghu2020eeg}, \emph{i.e.}, train a base network and then copy its first $n$ layers to those of a target network. A pre-trained model can leverage the knowledge gained from a large dataset to solve a different but similar task with a small dataset more effectively. The remaining layers of the target network are randomly initialized and trained towards the target task. One can either fine-tune the entire deep network, or freeze the first $n$ layers. This depends on the size of the target dataset and the number of parameters in the first $n$ layers. If the size of the target dataset is large or the number of parameters is small, it can be fine-tuned to the target domain to improve performance. If the size of the target dataset is small and the number of parameters is large, overfitting will likely occur, and hence the first $n$ layers are often frozen. For example, Raghu \emph{et al.} extracted features using pretrained network and used SVM for classifying the seizure type \cite{raghu2020eeg}, and it outperformed conventional feature and clustering based approaches. 

\begin{figure} [h]
	\centering
	\subfigure
	{
		\includegraphics[width=1\linewidth=1]{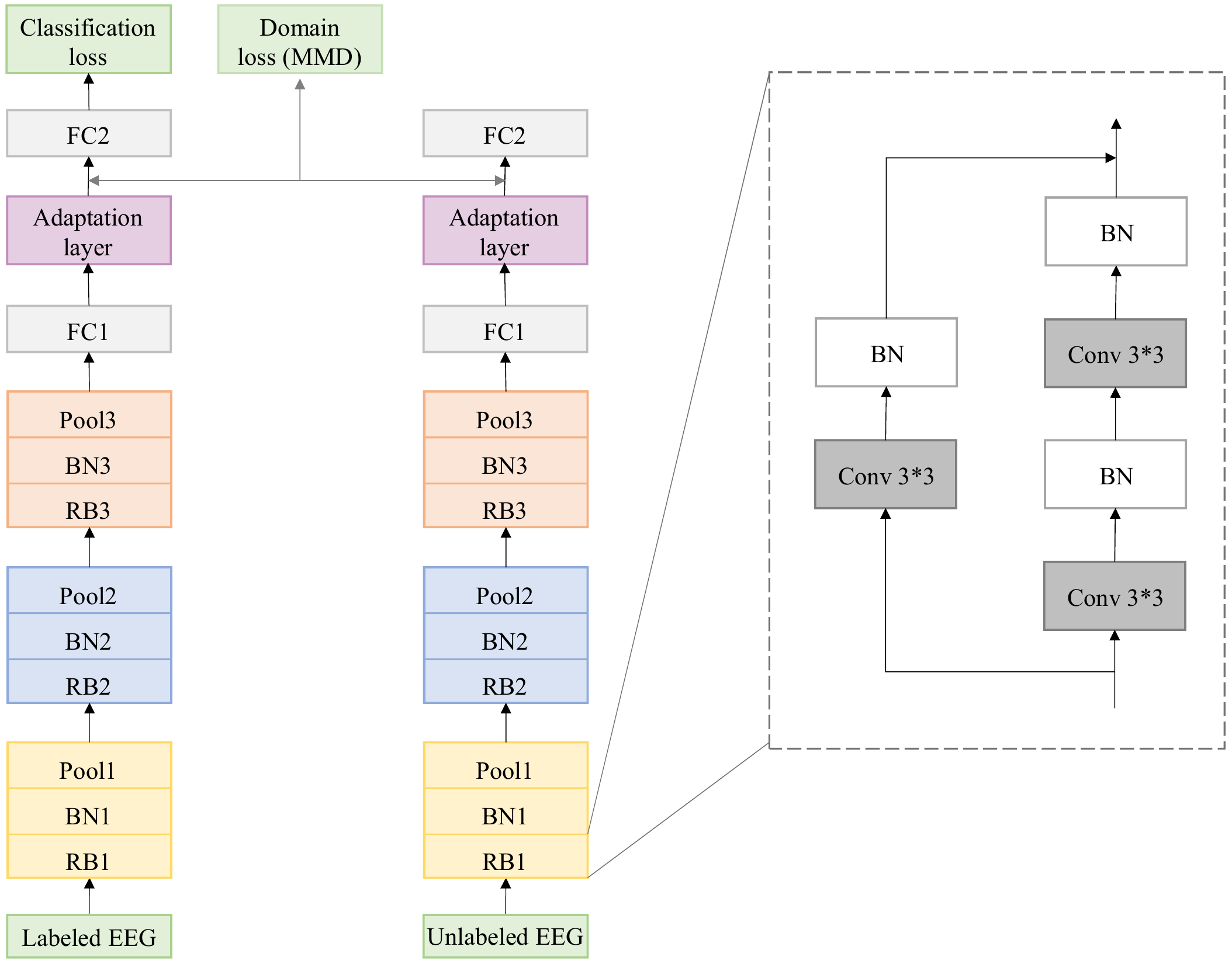}
	}
	\caption{Cross-subject recognition method based on CNN and DDC (Figure adapted from \cite{zhang2019cross}).}
	\label{TL2}
\end{figure}

Although fine-tuning is easy to implement and understand, it is less effective when there is substantial mismatch between the  distributions of source and target domains. To solve this problem, Zhang \emph{et al.} proposed a cross-subject recognition method based on a convolutional neural network (CNN) and deep domain confusion (DDC) \cite{zhang2019cross}, as shown in Fig. \ref{TL2}. They trained the CNN using the source and target EEG data jointly to minimize the loss of classification accuracy. The DDC method can narrow the difference in feature distribution between the source and target domains by minimizing the distance between the two domains (maximizing the domain confusion). Hence, a classifier trained in the source domain can be applied to the target domain to reduce the loss of classification accuracy. The maximum mean discrepancy (MMD) for maximizing domain confusion is defined as follows
\begin{equation}\label{Eq6}
\text{MMD}(X_S, X_T) = \bigg\| \frac{1}{|X_S|}\sum_{x_s\in{X_S}}\Phi(x_s) - \frac{1}{|X_T|}\sum_{x_t\in{X_T}}\Phi(x_t)\bigg \|_{\mathcal {H}}^2, 
\end{equation}
where $x_s$ and $x_t$ denote the deep features of the source and target domains in the adaptation layer, respectively. $|X_S|$ and $|X_T|$ denote the numbers of samples in the source and target domains, respectively. $\Phi$ denotes the kernel function that maps the deep feature to a reproducing kernel Hilbert space (RKHS). The total loss is defined as follows
\begin{equation}\label{Eq7}
L = L_C(X_L, Y_L) + \alpha \text{MMD}(X_S, X_T), 
\end{equation}
where $L_C(X_L, Y_L)$ denotes the classification loss in the tagged source domain EEG data $X_L$ and the corresponding ground truth labels $Y_L$, and $\alpha$ denotes the regularization parameter.

\begin{figure} [h]
	\centering
	\subfigure
	{
		\includegraphics[width=1\linewidth=1]{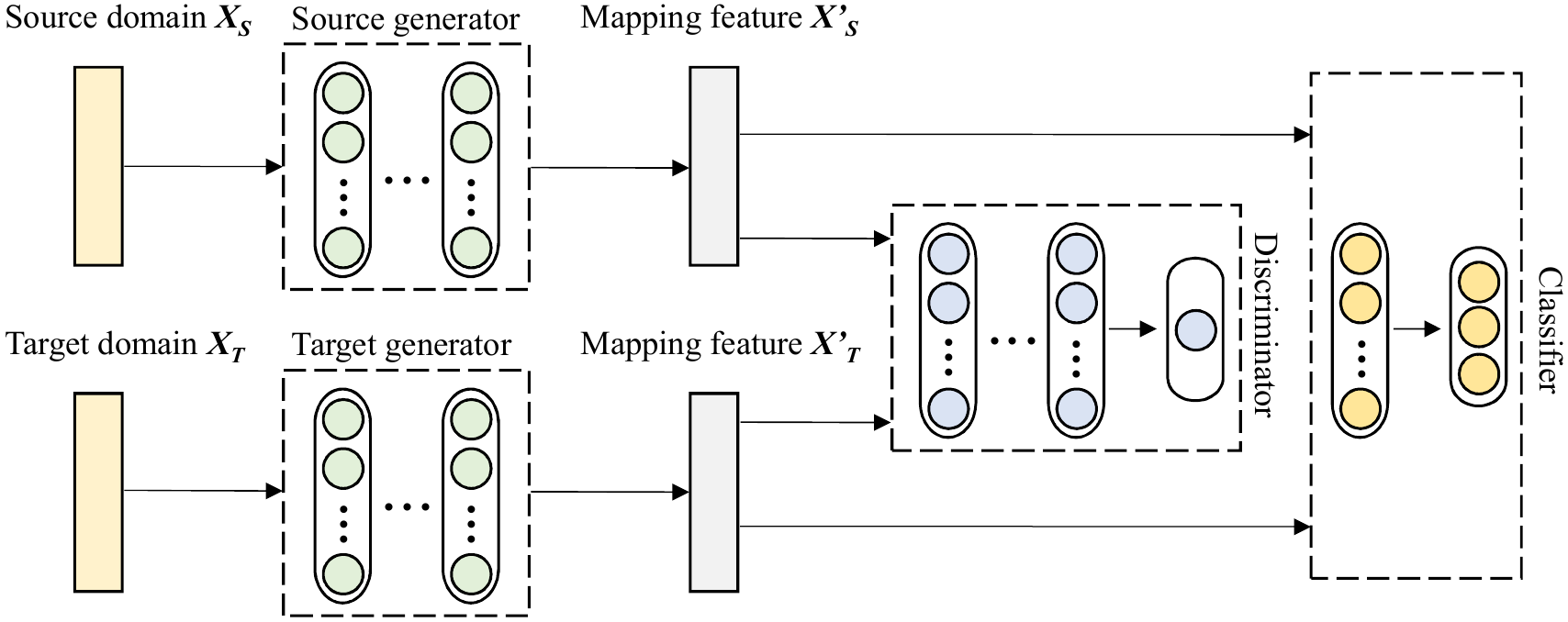}
	}
	\caption{Generative adversarial network domain adaptation framework (Figure adapted from \cite{luo2018wgan}).}
	\label{TL3}
\end{figure}

An alternative approach is to adopt generative adversarial domain adaptation \cite{luo2018wgan}, which is closely related to GANs. GANs have the advantage of generative ability and can be formulated as a minimax problem. The distribution of generated data is approximate to that of real data when the two-player game achieves equilibrium. Thus, as shown in Fig. \ref{TL3}, the generative adversarial domain adaptation \cite{luo2018wgan} framework has been widely used to solve the distribution mismatch problem in computer vision as well as in EEG decoding. The source and target generators aim to map the source and target domains to a common feature space, respectively. The discriminator aims to distinguish the source and target distributions in the common feature space, and the classifier is used to recognize the EEG state.

\begin{figure} [h]
	\centering
	\subfigure
	{
		\includegraphics[width=1\linewidth=1]{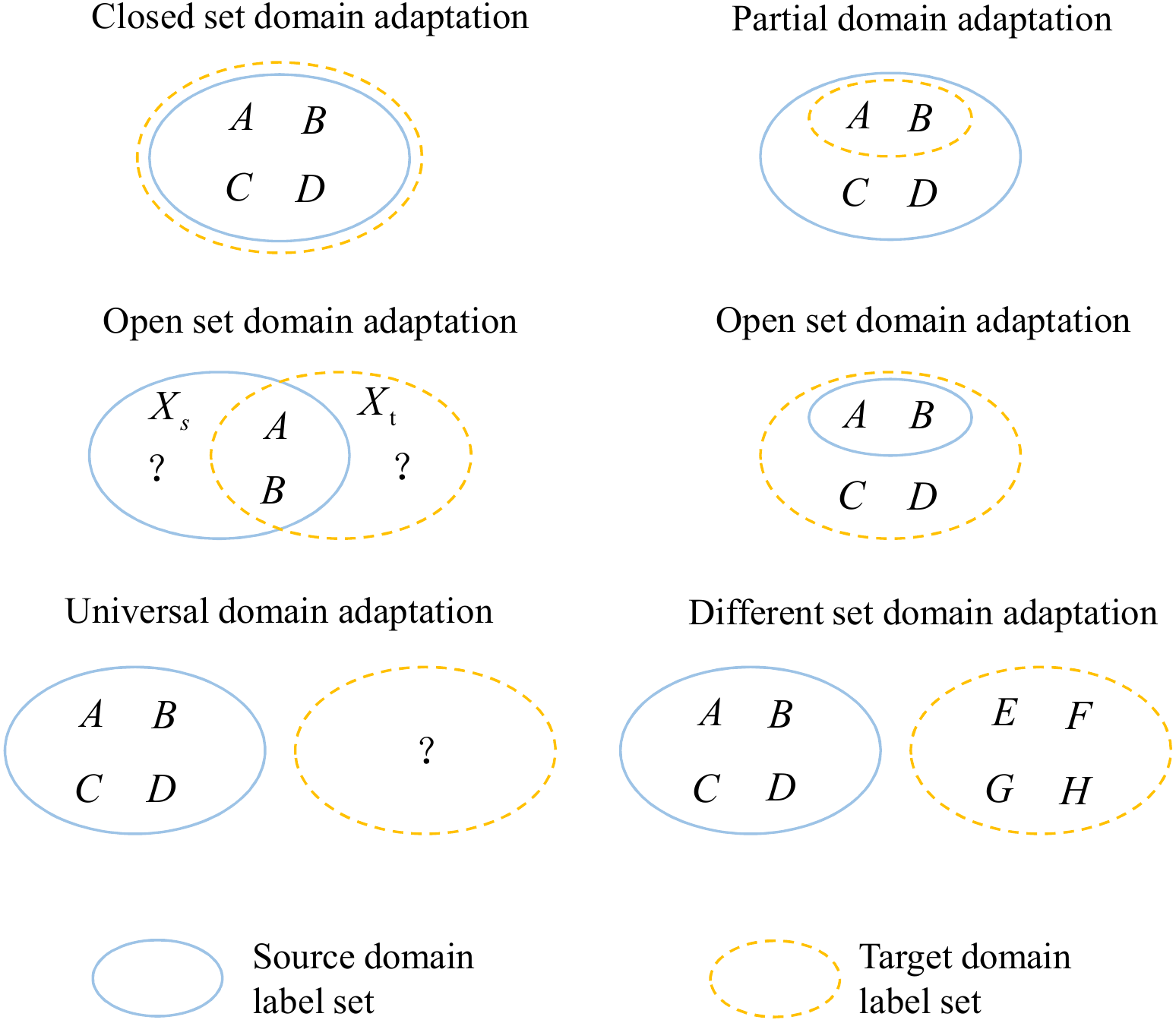}
	}
	\caption{Different domain adaptation scenarios \cite{he2020different}.}
	\label{TL4}
\end{figure}

Deep domain adaptation methods improve the model performance in the target domain by eliminating the domain shift between the source and target domains. However, most domain adaptation methods require the source domain to have the same feature space and label space as the target domain, which may not always be the case in open-world EEG-based applications. In fact, the source and target domains may have different label spaces. Figure \ref{TL4} shows different domain adaptation scenarios. In closed set domain adaptation, the source and target domains are assumed to have the same label space. In partial domain adaptation \cite{cao2018partial}, the classes of the target domain contain only a subset of the source domain, and in open set domain adaptation, it is assumed that the source and target domains have several common classes but also several classes that are different and unknown. In open set domain adaptation considered on the left \cite{panareda2017open} of the second row in Fig. \ref{TL4}, the source and target domains contain some common classes, but each also contains an ``unknown" class. In open set domain adaptation considered on the right \cite{saito2018open} of the second row in Fig. \ref{TL4}, the source domain only contains a subset of the target domain classes. In universal domain adaptation \cite{you2019universal}, the target domain may contain several common classes with the source domain. However, it may also contain several unknown classes. In different set domain adaptation, the target domain contains partially or completely different classes from the source domain. 

Domain adaptation methods require acquaintance with the target-domain data to measure the discrepancy between the source and target domains in the training stage. However, they require data collection and model training for each target domain (subject) that are high in cost and low in efficiency \cite{ma2019reducing}. Domain generalization aims to learn a model using data from a single or multiply-related but different source domains so that the model can generalize well to any target domain \cite{ma2019reducing}. For open-world EEG-based applications, domain generalization can extract domain-invariant features by exploiting domain differences among source subjects without access to the target subjects. Thus, domain generalization can be more robust in open-world applications when applied to unseen domains. For example, Ma \emph{et al.} proposed a domain generalization method by applying deep adversarial networks to reduce the influence of subject variability without requiring any information from unseen subjects, their method could generalize well to multiple test subjects compared with existing domain adaptation methods \cite{ma2019reducing}.

\subsubsection{Multi-task and multi-modal learning for exploring the joint and complementary information}
Some recent techniques are based on the observation that humans can learn multiple tasks simultaneously. They can use the knowledge learned in one task to help with learning of another, related task. The large numbers of annotated EEG samples typically required by DL methods for adequate recognition performance are almost impossible to obtain due to the high cost of data acquisition and accurate annotation. Multi-task learning is an approach that aims to improve the generalization performance of all tasks by leveraging useful information contained in multiple, related tasks \cite{zhang2018overview}. It is assumed that all tasks, or at least a subset of them, are related to each other. Learning multiple tasks jointly has theoretically and empirically been found to achieve better performance than learning them independently \cite{zhang2018overview}. Multi-task learning can allow access to more data overall, resulting in more robust and universal representations, and lower risk of overfitting for each task. Multi-task learning is related to transfer learning but has certain differences. In multi-task learning, all tasks are equal, and the aim is to improve the performance of all tasks \cite{ma2018predicting}. However, transfer learning aims to improve the performance of a target task with the help of source tasks. Thus, the target task attracts more attention than the source tasks.

\begin{figure} [h]
	\centering
	\subfigure
	{
		\includegraphics[width=0.9\linewidth=0.9]{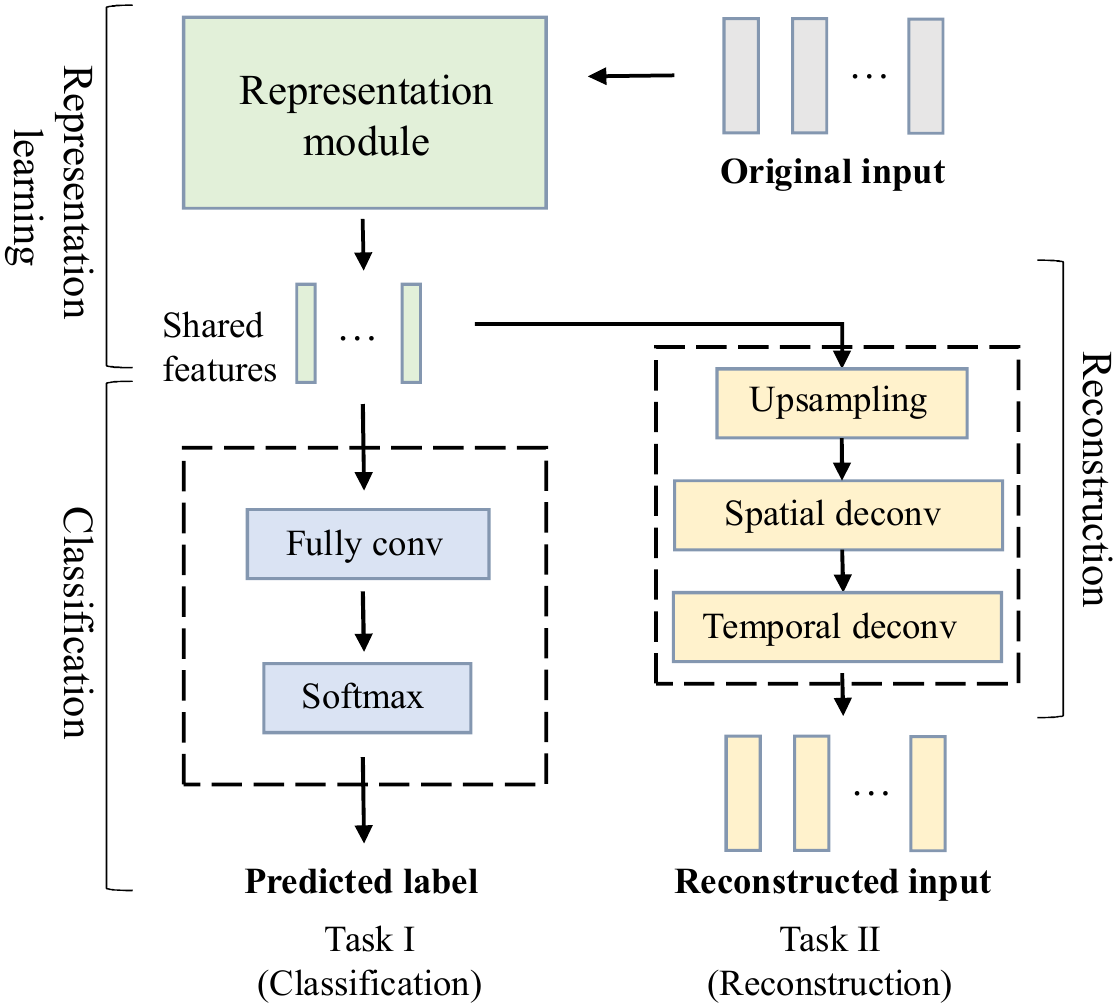}
	}
	\caption{EEG classification via multi-task DL (Figure adapted from \cite{song2019eeg}).}
	\label{MTL}
\end{figure}

Multi-task DL, where each task is solved by its own deep network, can improve performance over single-task DL if the associated tasks share complementary information or act as regularizers for one another \cite{ma2018predicting}. In addition,  the inherent layer-sharing representation can reduce the memory footprint and avoid calculating features repeatedly in the shared layers. Thus, it can yield fast learning speed and increase data efficiency for related or downstream tasks. For example, Song \emph{et al.} proposed an EEG classification method based on multi-task DL \cite{song2019eeg}, as shown in Fig. \ref{MTL}. It consists of three modules, one for each of representation, classification, and reconstruction. The representation module learns shared features from EEG signals which are then sent to the classification module for prediction and then the reconstruction module to reconstruct the original EEG signal. The shared features work as a bridge to unite the classification and reconstruction tasks, and the two tasks are jointly optimized in an end-to-end manner. Through the interaction of the two tasks, the shared features maintain both classification and reconstruction abilities. Therefore, it can enhance the generalization ability of the deep model and improve classification performance with limited EEG data. Abdon \emph{et al.} presented a multi-task cascaded deep neural network for joint prediction of people's affective factors using EEG signals recorded from people while watching affective videos in either individual or group configuration \cite{miranda2018multi}. The proposed network consists of two levels of prediction. The first level, affect network, is designed to predict the participant's affective levels of valence and arousal expressed by the participants during single video segments. The second level, personal factors network, uses the prediction of affective levels of consecutive video segments to perform multi-task prediction of personal factors. Liu \emph{et al.} proposed a multiscale space-time-frequency feature-guided multi-task learning convolutional neural network architecture for EEG classification \cite{liu2021multiscale}, which can fuse the complementary characteristics of different models. This method consists of four modules, \emph{i.e.}, the space-time feature-based representation module, time-frequency feature-based representation module, multi-modal fused feature-guided generation module, and classification module. The four modules are trained using three tasks simultaneously and jointly optimized. Due to the interaction of the three tasks, it can improve the generalization ability and accuracy of subject-dependent and subject-independent methods with limited annotated data.

Multi-modal learning aims to utilize complementary or supplementary information from different modalities to complete a shared task or multiple related tasks \cite{ramachandram2017deep}. The underlying motivation for using multi-modal data is that complementary or supplementary information can be extracted from different modalities for the shared task or multiple related tasks. It can obtain richer representation to achieve better performance than using only a single modality. Traditional multi-modal learning methods are shallow models that cannot learn the intrinsic representation of data. Thus, they cannot capture inter-modality representations and cross-modality complementary correlations of multi-modal data properly. However, DL models can learn a hierarchical representation of the data across hidden layers, and the learned representations of different modalities can be fused at various levels of abstraction.

Interestingly, multi-modal DL approaches appear to have some relevance for the visual pathway in the brain \cite{palazzo2020decoding}. Palazzo \emph{et al.} proposed a multi-modal DL method to learn a neural representation by classifying brain responses to natural images \cite{palazzo2020decoding}, and it can learn a joint brain-visual embedding and find similarities between brain representations and visual features. This embedding can be used to perform image classification, saliency detection, and visual scene analysis. The motivation is to learn reliable joint representations and find correspondences between visual and brain features that can decode brain representations. In turn, these representations can also be used to build better DL models. Jia \emph{et al.} proposed a multi-modal DL method for sleep stage classification by fusing EEG, EOG, and EMG signals \cite{jia2021sleepprintnet}. Separate data representations were designed for each signal type. These representations were then input into a deep model to extract features from EEG, EOG, and EMG signals, respectively. Finally, a feature fusion module was used to fuse all extracted features for sleep stage classification. Cai \emph{et al.} proposed a feature-level fusion method based on multi-modal EEG data for depression recognition \cite{cai2020feature}. The multi-modal EEG data were acquired under neutral, negative and positive audio stimulation to discriminate between depressed patients and normal controls. Then, a feature-level fusion method was used to fuse the EEG data of different modalities  to construct a depression recognition model.

\subsubsection{Adversarial training for augmenting the training set with adversarial examples}
\begin{figure} [h]
	\centering
	\subfigure
	{
		\includegraphics[width=1\linewidth=1]{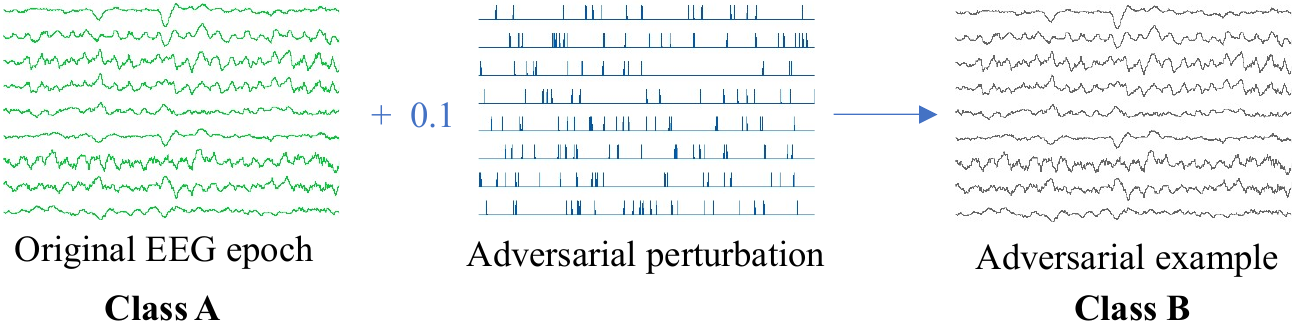}
	}
	\caption{A normal EEG epoch and its adversarial example.}
	\label{AL}
\end{figure}

Although DL models have achieved outstanding performance in EEG feature extraction, they are vulnerable to adversarial attacks, where normal EEG samples are corrupted with small, seemingly innocuous perturbations \cite{zhang2019vulnerability}. DL models have been found to be easily fooled by adversarial examples, which are normal EEG examples with small perturbations. Figure \ref{AL} shows a normal EEG epoch and a corresponding adversarial example. The perturbations are usually too small for the human eyes to perceive. However, despite the slight changes, adversarial EEG samples can lead to a dramatic performance degradation with possible serious consequences. For example, adversarial attacks could lead to misdiagnosis of disorders of consciousness in patients in clinical applications. Adversarial perturbations can mislead the P300 and steady-state visual evoked potential brain--computer interface (BCI) spellers to spell anything the attacker wants \cite{wu2021adversarial}. Modern EEG monitoring systems designed to detect epileptic seizures could be vulnerable to an adversarial attach whereby an ictal (seizure) sample would be classified as inter-ictal (non-seizure) in an emergency situations, with possible dire implications \cite{wu2021adversarial}.

Adversarial attacks can be classified according to the degree of access the attacker has to the target model  \cite{liu2021universal}, \emph{i.e.}, white-box attacks, black-box attacks, and gray-box attacks. White-box attacks \cite{miller2020adversarial} assume that the attacker can obtain all information of both the classifier and the defense mechanism. Black-box attacks assume that the attacker does not know the architecture or parameters of the target model. However, they can obtain the response to the input. Gray-box attacks assume that the attacker can obtain some information of the target model.  Adversarial attack can also be classified according to the stage that the attack is performed, including poisoning attacks and evasion attacks \cite{wu2021adversarial}. A poisoning attack takes place during the training time of the machine learning model \cite{chakraborty2018adversarial}. An adversary tries to poison the training data by injecting carefully-designed samples to eventually compromise the whole learning process. An evasion attack, the most common type, tries to evade the system by adjusting malicious samples during the testing phase \cite{chakraborty2018adversarial}. This setting does not assume any influence on the training data. According to the outcome, there are two types of adversarial attacks \cite{wu2021adversarial}, namely, targeted attacks and non-targeted (indiscriminate) attacks. Targeted attacks force a model to classify either a particular subset of data samples or a particular region of feature space to a chosen (usually wrong) class. Non-targeted attacks force a model to misclassify certain data samples or regions of feature space, but do not specify which class they should be misclassified into. 

In an open-world environment, obtained EEG signals are sent to a computer, a smart phone, or the cloud for further analysis. Goodfellow \emph{et al.} proposed the fast gradient sign method (FGSM) \cite{goodfellow2014explaining}, and soon it became a benchmark attack approach. Let $g$ be the deep learning model, $\theta$ be its parameters, and $J$ be the loss function for training $g$. The main idea of FGSM is to find an optimal perturbation $\eta$ constrained by $\varepsilon$ to maximize $J$. The perturbation can be calculated as follows
\begin{equation}\label{Eq8}
 \eta = \varepsilon \cdot \text{sign}(\bigtriangledown_{x_i} J(\theta,x_i,y_i)),
\end{equation}
where $x_i $ and $y_i $ represent the $i$th EEG trial and the corresponding label, respectively. It is not enough to know the architecture and parameters $\theta$ of the target model $g$, since it needs to know the true label $y_i $ of $x_i $ to generate the adversarial perturbation. Liu \emph{et al.} propose an unsupervised FGSM (UFGSM) to deal with this problem \cite{zhang2019vulnerability}. UFGSM replaces the label $y_i $ by $y_i' = g(x_i) $ , \emph{i.e.}, the estimated label from the deep model. The perturbation in UFGSM can be rewritten as follows
\begin{equation}\label{Eq9}
 \eta = \varepsilon \cdot \text{sign}(\bigtriangledown_{x_i} J(\theta,x_i,y_i')).
\end{equation}
All EEG trials are needed to determine an adversarial perturbation for each trial. Here, EEG trial means the EEG signal corresponding to a trial. For example, during each trial in motor imagery (MI), the subject is required to perform either of the two (right hand and right foot) MI tasks for 3.5-$s$. However, it is inconvenient to compute the adversarial perturbation for each EEG trial.  In addition, it requires all the EEG trials in advance to compute the adversarial perturbation \cite{liu2021universal}. It is impossible to attack as soon as an EEG trial starts. To address these issues, Liu \emph{et al.} introduced a universal adversarial perturbation \cite{moosavi2017universal} method that could obtain the universal perturbation template offline and hence attack open-world EEG-based systems in real time. Therefore, it is critically important to pay attention to security concerns of open-world EEG-based systems. Meng \emph{et al.} performed poisoning attack of EEG-based BCIs \cite{meng2020eeg}, and they proposed a practically realizable backdoor key, which can be inserted into original EEG signals during data acquisition.

To deal with adversarial attacks in EEG-based systems, many adversarial defense methods have been proposed \cite{wu2021adversarial}. The most representative method is to augment DL models with adversarial training \cite{goodfellow2014explaining}. Hussein \emph{et al.} proposed a method to augment DL models with adversarial training for robust prediction of epilepsy seizures \cite{hussein2020augmenting}.  First, a DL classifier is constructed from available limited amount of labeled EEG data, and adversarial examples are obtained by performing white-box on the classifier. Then, the training set is augmented with adversarial examples. Finally, DL models are retrained with the augmented training set, which can improve the robustness of DL models in open-world EEG-based systems. 

\subsection{Classification}

\subsubsection{Few-shot, zero-shot and semi-supervised learning for small sample size}
A variety of DL methods have shown superior performance compared to traditional methods in EEG classification \cite{roy2019deep}. For example, Tabar \emph{et al.} proposed CNN and stacked autoencoders (SAE) to classify EEG MI signals \cite{tabar2016novel}, which can obtain better classification performance compared with other traditional methods. However, with inadequate training samples, DL models are prone to overfitting, which leads to a decrease in classification accuracy. Although each trial can be over-sampled to obtain a larger number of samples, these samples are highly dependent on each other, so that a larger number of EEG trials are still preferable to achieve reliable performance. BCI feedback applications typically require a tedious calibration process that can be challenging in some patient populations \cite{shim2020gradual}. Clearly designing robust DL models for small sample sizes is important in EEG classification \cite{wang2020generalizing} \cite{keshari2020unravelling}.

To learn from a limited number of EEG training examples with supervised information, Cheng \emph{et al.} proposed a deep forest model named multi-Grained Cascade Forest (gcForest) for multi-channel EEG-based emotion recognition task \cite{cheng2020emotion}. This method is insensitive to hyper-parameter setting, and greatly reduces the complexity of EEG emotion recognition. The model complexity of gcForest can be determined automatically for different size of training data, making it suitable for small-scale training data. In addition, a new machine learning paradigm called few-shot learning has been proposed \cite{wang2020generalizing}. The goal of few-shot learning is to classify unseen data instances (query data) into a set of classes, given just a small number of labeled instances (support examples) in each class. Typically, there are between 1 and 10 labeled support examples per class in the support set. 

The EEG recording during MI (used to aid rehabilitation as well as autonomous driving) is a good scenario where these issues arise and has driven the development in this area. MI also allows users to generate the suppression of oscillatory neural activity in specific frequency bands over the motor cortex region without external stimuli \cite{pfurtscheller1999event}. The neurophysiological patterns of MI originate from changing brain areas' activations in the sensorimotor cortices similar to limb movements. Furthermore, a recent study has demonstrated MI-based BCI as an assistive tool in post-stroke motor rehabilitation. However, due to the scarcity of unseen subject data, complex dynamics of MI signals, inter-subject variability, and low signal-to-noise ratio, it is still challenging to improve the performance of MI-based classification tasks. 

\begin{figure} [h]
	\centering
	\subfigure
	{
		\includegraphics[width=1\linewidth=1]{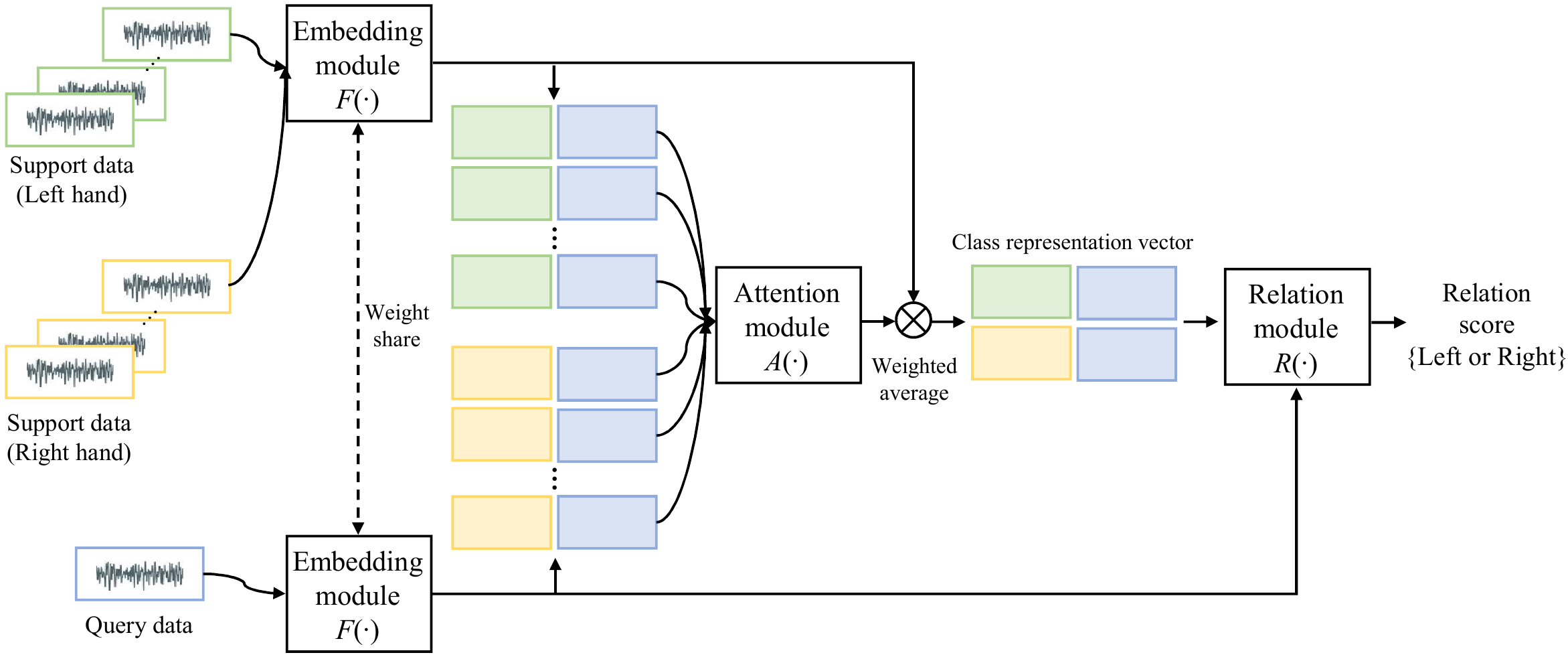}
	}
	\caption{Few-shot relation network (Figure adapted from \cite{an2020few}).}
	\label{FSL}
\end{figure}

To solve the above mentioned problems, An \emph{et al.} formulated an EEG-based MI classification task as a few-shot learning problem \cite{an2020few} (Fig. \ref{FSL}), and it could classify unseen subject data with a small number of MI EEG data. They also proposed a novel few-shot relation network consisting of a feature embedding module, attention module, and relation module in an end-to-end framework. The embedding module is used to extract semantic features from the support and query data. Given the extracted semantic features, the attention module is used to obtain the attention score for each support sample using both support and query features. Then, the representative vector for each class can be obtained using a weighted average of $k$ support features with attention scores. Finally, the relation module is used to obtain the relation scores based on the distance metric between class-representative vectors and the query features. During training, the few-shot relation network is trained using pairs of support set and query data among different subjects in the training data. During testing, the label of a query datum can be taken as the class with the largest predicted relation score by using the $k$ labeled support signals from an unseen subject. Therefore, the few-shot relation network enables good generalization ability to classify the query data of an unseen subject, even with a small amount of MI EEG data.

MI usually has only a few mental tasks that can be translated to corresponding commands (\emph{e.g.}, imagine left-hand or right-hand movements). In a real-time BCI environment, a calibration procedure is particularly necessary for each user and each session. The aim of the calibration is to derive individualized parameterization of EEG signals. In fact, to construct the classifier, a separate parameterization is sought for discriminating each intentional control (IC) state, related to different MI, from the noncontrol (NC) state, and for discriminating among various IC states. The signal processing procedures of calibration include obtaining individualized parameters for classification, updating these parameters after the subsequent user training, and online signal processing and classification for BCI operation. The calibration consumes a significant amount of time that hinders the application of a BCI system in a real open-world scenario \cite{shim2020gradual}. For example, the MI system requires a considerable amount of time to record sufficient EEG data for robust classifiers' training. Owing to these inevitable BCI environments, the MI-BCI system has considered the brain dynamics, reflecting each individual's EEG characteristics. In addition, the subjects tend toward a state of inattention state in real-time experiments due to the long calibration times for offline experiments \cite{jiao2018sparse}. Despite the improved performance over the conventional methods, the deep learning methods often fail when training samples per subject are limited. Thus, a huge number of	training samples need to be obtained from each target subject to train the robust model \cite{an2020few}. Moreover, the MI-based BCI system is often limited by the types of MI \cite{duan2020zero}. Usually, only a few mental tasks, such as the movements of left-hand, right-hand, and foot, can be recognized and translated to corresponding commands. Thus, it is important to simultaneously reduce the calibration time and increase the number of commands. 

The learned classifier in supervised classification can only recognize the categories of target EEG signals that have been seen during training. However, it cannot deal with previously unseen classes. Seen or known classes refer to classes that are covered by the training dataset, while unseen or unknown classes refer to those that were not seen in the training dataset. Zero-shot learning, a powerful and promising learning paradigm \cite{wang2019survey}, can recognize unknown categories of EEG signals \cite{duan2020zero}, and has the potential to substantially reduce the calibration time. In zero-shot learning, there are some labeled training instances in the feature space. The classes covered by these training instances are referred to the known classes or seen classes. There are also some testing instances that do not belong to any known classes. These classes are referred to as the unknown classes or unseen classes. It is assumed that the classes covered by the training dataset and the classes of the testing dataset are disjoint. 

\begin{figure} [h]
	\centering
	\subfigure
	{
		\includegraphics[width=1\linewidth=1]{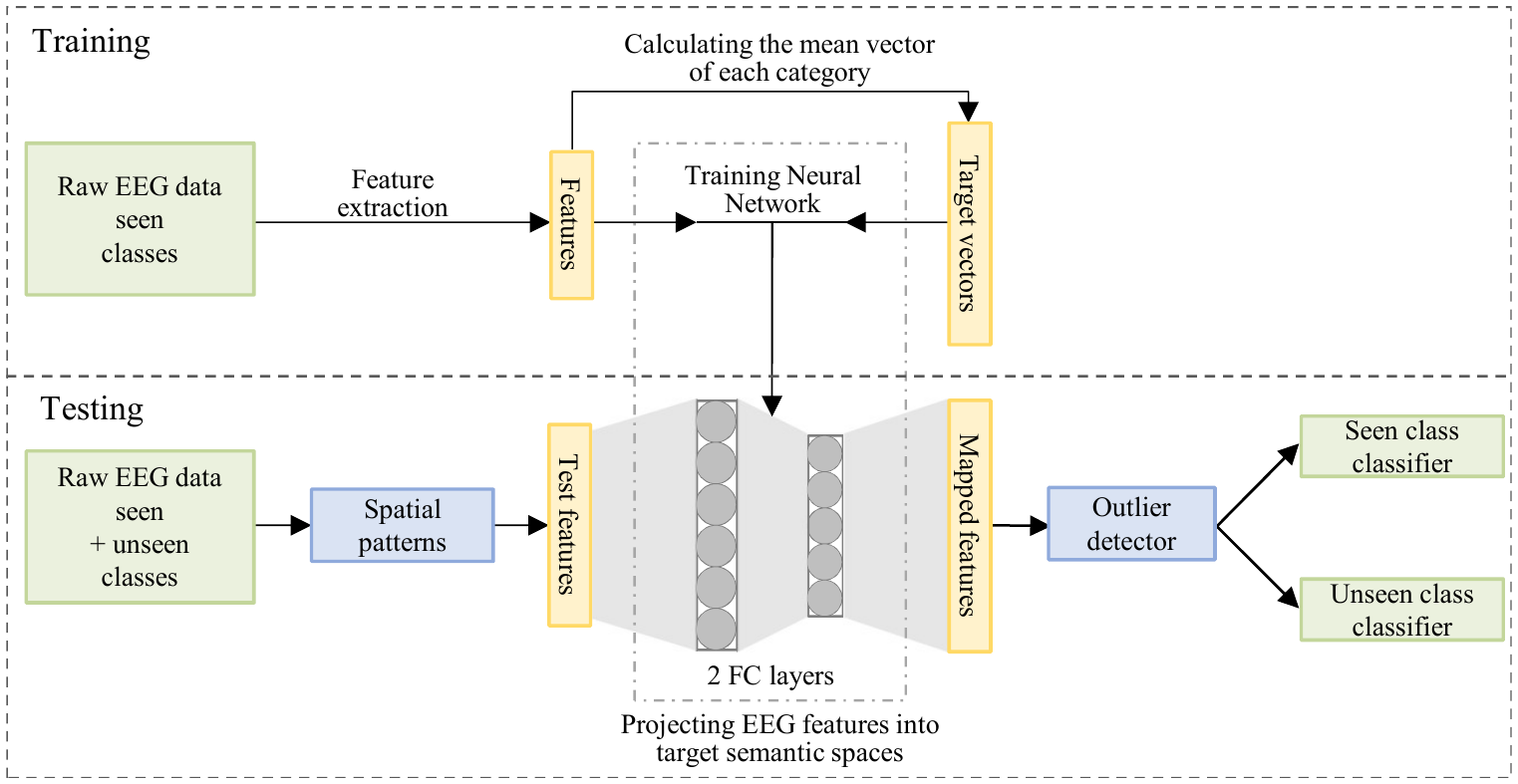}
	}
	\caption{Zero-shot learning for EEG based MI classification (Figure adapted from \cite{duan2020zero}).}
	\label{ZSL}
\end{figure}

However, the assumptions of zero-shot learning are so restrictive that it can only predict unknown classes. Thus, generalized zero-shot learning has been proposed to recognize both known and unknown categories of EEG signals. Duan \emph{et al.} proposed a generalized zero-shot learning method for EEG classification in a MI-based BCI system \cite{duan2020zero}, as shown in Fig. \ref{ZSL}. It could recognize unknown task samples (\emph{e.g.}, imagine left and right hands move simultaneously). The first step is to extract features from EEG signals using a common spatial pattern, and it aims to obtain discriminative spatial patterns by maximizing the variance ratios of filtered EEG signals for two classes. Then, the obtained EEG feature vectors are projected onto the target semantic space, which can help to recognize unknown task samples. The mean vector of each class is taken as semantic information, which can capture the distribution of different classes. Two fully connected layers with a \emph{tanh} activation function are used to map all the samples in each class to the corresponding mean vector. In this manner, testing samples of the known classes are clustered around the training samples from two classes (left-hand and right-hand MI). However, the unknown task samples are far away from the known classes. An outlier detection method is used to determine whether a mapped EEG feature belongs to known classes. If the mapped feature belongs to a known class, a classifier can be used to determine the class. Otherwise, it is assigned to a class based on the likelihood of being an unknown class. Therefore, the generalized zero-shot learning method can recognize not only unknown classes but also known classes. Hwang \emph{et al.} proposed a new framework for zero-shot EEG signal classification \cite{hwang2019ezsl}, which has three parts. The first part is an EEG encoder network that generates EEG features. The second part is a GAN that can recognize the unknown EEG labels with a knowledge base. The third part is a simple classification network to learn unseen EEG signals from the fake EEG features that are generated from the learned generator.

To solve the small sample size problem in EEG classification, semi-supervised learning adopts a small number of labeled EEG data and a large amount of unlabeled EEG data simultaneously, which can be seen as a combination of supervised and unsupervised learning \cite{van2020survey}. Since a large amount of unlabeled EEG data can help to capture the underlying distribution of data, semi-supervised learning methods can improve the performance of supervised learning by using unlabeled EEG data to learn more robust representations and hence alleviate the need for large amounts of labeled EEG data. For example, Jia \emph{et al.} proposed a novel, semi-supervised DL framework for EEG emotion recognition \cite{jia2014novel}. They used label information in feature extraction and integrated the unlabeled information to regularize the supervised training. Specifically, they determined a sample's potential contribution to the model training based on the uncertainty of the trained model over each unlabeled EEG sample. After the training stage, it can quickly predict the label for a test sample. With careful scrutinization, it is observed that the result is sometimes unreliable, that is, the value of conditional probability $P(y|X)$ for a test sample $X$ with different label $y$ can be very close. On the contrary, if the probability value for certain label $y$ dominates the others, the model is quite confident with its decision. In this way, it can conclude that the sample in the former case contains more uncertainty \cite{jia2014novel}, and it can result in a faster advance to the more accurate decision boundary. Thus, both supervised and unsupervised information can be jointly utilized in the entire training process to reduce model variance. Panwar \emph{et al.} proposed a semi-supervised Wasserstein GAN with gradient penalty to classify driving fatigue from EEG signals \cite{panwar2019semi}. This method is an extension of the Wasserstein GAN to include a classifier that predicts the class labels of the data, which enables the augmentation of limited training samples with generated EEG samples during training, and hence leads to improved classification performance.
 
\subsubsection{Self-supervised learning for discovering structure in unlabeled EEG data}
Most EEG-based DL models utilize supervised learning methods. However, supervised learning methods have several limitations. It is well known that DL requires a large amount of labeled EEG data to achieve satisfactory performance. However, most EEG studies are conducted under small labeled data regimes, and a few hundred subjects are often considered as big data. Large-scale EEG-based supervised learning is much rarer. Thus, it is difficult to achieve a desirable performance for most EEG-based DL models. In addition, supervised DL models must be trained from scratch for each task, and they require a large amount of computational resources and time. The learned representations are often very task-specific and are not expected to generalize well to other tasks. Furthermore, it is challenging to know exactly what the participants are thinking or doing in cognitive neuroscience experiments, and hence, it is difficult to obtain accurate labels \cite{banville2020uncovering}. Finally, supervised EEG-based DL models are prone to adversarial attacks.

Self-supervised learning, an unsupervised learning paradigm, can learn the underlying features from large-scale unlabeled data without using any labeled data, thereby avoiding the extensive cost of collecting and annotating large-scale datasets. Self-supervised learning usually reformulates the unsupervised learning problem as a supervised learning problem and is composed of a pretext task and downstream task \cite{banville2020uncovering}. Pretext tasks are pre-designed tasks that are helpful for downstream tasks. The supervision signal is generated from the data itself by leveraging the structure, instead of manual annotation, and the features can be learned by training the objective function of pretext tasks. After training the pretext task, the features learned by the pretext task can be transferred to the downstream task. By training the model to solve well-designed pretext tasks, self-supervised learning can help the model to learn more generalized representations from unlabeled data. Thus, it can achieve better performance and generalization on downstream tasks. In general, annotated labels are required to solve downstream tasks. However, in several applications, the downstream task can be the same as the pretext task without using any annotated labels.

Self-supervised learning has several advantages over supervised learning for EEGs. For example, self-supervised learning can exploit the structure of unlabeled data to provide supervision. The learned representations are often more general than task-specific supervised learning, and more robust to inter-class variation and intra-class variation. Thus, it can be reused for different tasks and save computation time compared to training a model from scratch for each task. In addition, self-supervised learning can make full use of large amounts of unlabeled data, which in turn can help to train much deeper and more sophisticated networks. Self-supervised learning can also improve the performance of DL when there are limited or no labeled data. Banville \emph{et al.} investigated self-supervised learning to determine the structure in unlabeled data to learn useful representations of EEG signals \cite{banville2020uncovering}. Three pretext tasks were designed for EEG, \emph{i.e.}, relative positioning, temporal shuffling, and contrastive predictive coding. Relative positioning aims to discriminate pairs of EEG samples based on their relative positions. This is based on the assumption that an appropriate representation of the data should evolve slowly over time and EEG samples close in time should share the same label. Pairs of windows are sampled from time series $S$ (EEG recording) so that the two windows of a pair are either close in time (`positive pairs') or farther away (`negative pairs'), and an end-to-end feature extractor $h_\varTheta$ is trained to predict whether a pair is positive or negative. For temporal shuffling, triplets of windows are sampled from $S$, and a triplet is given a positive label if its windows are ordered or a negative label if they are shuffled. For contrastive predictive coding, sequences of $N_c + N_p$ consecutive windows are sampled from $S$ along with random distractor windows (`negative samples'). Given the first $N_c$ windows of a sequence (`context'), a neural network is trained to identify which window out of a set of distractor windows actually follows the context. Downstream tasks were performed on two EEG-based clinical applications, sleep staging and pathology detection. Experiments demonstrated that linear classifiers trained on self-supervised learned features can outperform supervised deep models under small labeled data regimes \cite{banville2020uncovering} and achieve competitive performance when all labels are available. In addition, the learned representation can reveal the latent structures related to physiological and clinical phenomena. Furthermore, all self-supervised learning tasks systematically outperformed or matched the compared methods in low-to-medium labeled data regimes, and remained competitive in a high labeled data regime.

Recently, contrastive learning has been successful in computer vision for representation learning. Chen \emph{et al.} introduced a simple framework for contrastive learning of visual representations (SimCLR) \cite{chen2020simple}, which can learn representations that are invariant under a set of augmentations through a contrastive loss. To learn EEG representations, Mohsenvand \emph{et al.} modify the SimCLR framework to work with time-series EEG data \cite{mohsenvand2020contrastive}. The core idea of contrastive self-supervised learning is to train the networks while maximizing similarity for augmented instances of the same data point and minimizing the similarity between different data points. In contrast to images where the set of augmentations are intuitive and easily verifiable by the human eye, it is not clear what augmentations could be beneficial for EEG. They trained a channel-wise feature extractor by extending the contrastive learning framework to time-series data and introduced a set of augmentations for EEG. Mohsenvand \emph{et al.} consulted four neurologists and two postdoctoral researchers at anonymized hospital research group specializing in clinical interpretation of EEG to identify a set of augmentations that do not change the interpretation of EEG data \cite{mohsenvand2020contrastive}. They chose the transformations that were easy to randomize programmatically, and ran preliminary experiments to choose a minimal effective set.  Experiments demonstrated that the learned features can improve the accuracy of EEG classification and significantly reduce the amount of labeled data required for three EEG tasks, \emph{i.e.}, emotion recognition, normal/abnormal EEG classification, and sleep-stage scoring.

\subsubsection{Robust EEG classification in the presence of noisy label}
In EEG experiments, each trial is associated with a stimulus and response, \emph{i.e.}, the stimulus to the participant and the behavioral response of the participant to it. The behavioral response can be a label, and it is assumed to be in accordance with the stimulus \cite{porbadnigk2014brain}. However, it requires a large amount of well-labeled EEG data for deep supervised learning to achieve desirable performance, and this may not always be available for real open-world applications. It is difficult to interpret and annotate EEG signals due to noise in the data and the complexity of brain processes, which can lead to high inter-rater variability, \emph{i.e.}, label noise \cite{banville2020uncovering}. In addition, poisoning attacks can poison the training data by modifying their labels \cite{biggio2011support}. Label noise: the sample is valid but the label is wrong due to mislabeling. Data noise: the data is noisy but the label is valid, for example, samples caused by corruption, occlusion, distortion, and so on. Also, it is challenging to know exactly what the participants are thinking or doing in cognitive neuroscience experiments. In imagery tasks, for instance, the subjects might not be following instructions or the process under study might be difficult to quantify objectively (\emph{e.g.} meditation, emotions). For example, participants may not always generate the intended emotions when watching emotion-eliciting stimuli \cite{zhong2020eeg}. Moreover, if participants become sleepy, bored, or distracted \cite{porbadnigk2014brain}, it would lead to a significant increase in mislabeled trials. DL models would overfit to noisy labels due to the capability of learning any complex function, and the parameters obtained after training would deviate from the true optimal value, leading to a decline in the classification accuracy during testing. Therefore, it is necessary to consider noisy labels in real open-world applications. For example, in the field of EEG-based emotion recognition, humans have natural bias and inconsistencies in their judgments, which creates noise in their ratings. It is generally acknowledged that emotions are subjective, and studies have indicated that humans understand and perceive emotions varyingly. Moreover, participants may not always generate the intended emotion when watching emotion-eliciting stimuli, and the emotion label may be noisy and inconsistent with the actual elicited emotions. Zhong \emph{et al.} proposed a regularized graph neural network for EEG emotion recognition using node-wise domain adversarial training and emotion-aware distribution learning to deal with noisy labels \cite{zhong2020eeg}. Instead of learning the traditional single-label classification, the emotion-aware distribution learning method learns a distribution of labels of the training data and thus acts as a regularizer to improve the robustness of our model against noisy labels. Up to now, few works have been proposed for EEG classification in the presence of noisy label. A number of methods have been proposed for image classification with DL in the presence of noisy labels \cite{algan2021image}, which can provide inspiration for noisy label EEG classification. For example, Patrini \emph{et al.} proposed a loss correction method to make deep neural networks robust to label noise \cite{patrini2017making}, and the minimizer of the corrected loss under the noisy distribution was the same as the minimizer of the original loss under the clean distribution. Huang \emph{et al.} proposed a simple but effective noisy label detection method for deep neural networks without human annotations \cite{huang2019o2u}, and it only required adjusting the hyper-parameters of the deep neural network to make it transfer from overfitting to underfitting cyclically. Ren \emph{et al.} proposed a novel method that learns to assign weights to training examples based on their gradient directions \cite{ren2018learning}, and it adopted a meta gradient descent step on the current mini-batch example weights to minimize the loss on a clean unbiased validation set.
 
\begin{table*} \renewcommand{\arraystretch}{1}
	\caption{Challenges and future directions. }
	\label{table2}
	\centering
	\begin{tabular}{*{3}{|c}|}
		\hline
		& Challenges & Possible solutions  \\
		\hline
		\multirow{2}{*}{Preprocessing} & How to solve the domain shift between real-world EEG noise and simulated data noise & Designing a more universal deep network\\
		 \cline{2-3}
		 & How to design deep models without access to paired clean-noisy EEG samples & Unpaired deep models via GAN \\
		 \hline
		 \multirow{2}{*}{Feature extraction} & How to perform domain adaptation without access to the source EEG data &  Source-free unsupervised domain adaptation\\
		 \cline{2-3}
		 & How to exploit multiple source and target EEG data in transfer learning & Multiple source and target domain adaptation \\
		 \hline
		 \multirow{7}{*}{Classification} & How to solve the class imbalance problem in EEG signal analysis & Imbalanced learning \\
		 \cline{2-3}
		 & How to train a deep model incrementally from a stream of EEG samples & Online learning \\
		  \cline{2-3}
		 & How to design the architecture of a deep network automatically for EEG recognition & Neural architecture search \\
		  \cline{2-3}
		 & How to deploy deep models on portable EEG devices with limited resources & Model compression \\
		  \cline{2-3}
		 & How to achieve a higher level of automation in solving diverse EEG-based tasks & Automatic machine learning \\
		  \cline{2-3}
		 & How to exploit the interpretability of deep models for EEG analysis & Interpretable DL \\
		  \cline{2-3}
		 & How to design EEG models that are comparable or superior to human intelligence & Strong artificial intelligence methods \\
		  \hline
	\end{tabular}
\end{table*}

\section{Conclusion and future directions}
EEG decoding has made significant progress in the past few decades. In this article, we comprehensively surveyed existing EEG decoding methods using DL in the open-world setting. The challenges facing open-world EEG decoding were described for each step involved, \emph{i.e.}, preprocessing, feature extraction and classification. In addition, approaches for solving open-world EEG decoding using DL were briefly introduced and summarized according to the core concepts, theory, progress, and examples. Despite the considerable progress that has been achieved in open-world EEG decoding using DL, there are several problems to be solved in future work, as illustrated in Table \ref{table2}.

The data noise in real applications is much more complex and with a different distribution than simulated EEG data noise. The characteristics of real-world data noise can differ with regard to different EEG collection settings and conditions. Thus, the problem of domain shift between real-world noise and simulated data noise should be considered in open-world EEG artifact removal. Traditional DL-based artifact removal methods aim to learn an end-to-end nonlinear function to map a noisy EEG signal to a clean EEG signal with known statistics. However, they tend to lack flexibility for blind and real open-world data noise. For example, a traditional deep network is trained under a specific level of SNR, and the trained network would fail for an unseen noise level. Thus, a more universal network should be trained with the entire expected range of SNR for blind denoising of contaminated EEG signals. The denoising performance of most CNN-based methods largely relies on supervised learning with a large amount of paired clean-noisy EEG signals. However, it would be difficult to collect true clean EEG signals for several open-world EEG decoding applications, and new DL-based artifact removal methods should be designed without access to clean EEG training examples or to paired clean-noisy EEG training examples. Unpaired DL methods can be used to solve this problem. In addition, it is important to train DL-based artifact removal methods using only noisy EEG signals. In sum, EEG artifact removal in open-world EEG decoding is a highly challenging task. Data noise can be natural or added by an adversarial attack. Similarly, label noise can be natural, or added by a poisoning attack. When there are both nature noise and adversarial noise for data and label, this is the worst-case for data noise and label noise in open-world EEG decoding, which needs to delve deep research in future work.

Most traditional domain adaptation methods in EEG signal analysis are assumed to have access to the source data during training. In several open-world EEG decoding applications, the requirements for accessing source domain data are restrictive. Sharing data can be a concern due to privacy and security issues. In addition, it is difficult to store, transmit, and process a large amount of source EEG data. Thus, it is necessary to conduct source-free unsupervised domain adaptation, where the pre-trained model of the source domain is expected to adapt to unlabeled target data. Moreover, unsupervised domain adaptation models are usually applied to a single source domain and a single target domain, while multi-source and multi-target domain adaptation are typically encountered in open-world EEG decoding. Thus, single-domain adaptation may be suboptimal as it ignores the knowledge shared across multiple domains. Furthermore, when applying the classifier trained on one dataset to other datasets, the performance will be degraded significantly, and it is important to improve both cross-subject and cross-dataset classification performance.  

In terms of other solutions for solving open-world EEG decoding using DL, only a few studies have started to focus on few-shot learning, zero-shot learning, semi-supervised learning, self-supervised learning, and noisy labels for EEG analysis. Thus, there is considerable scope for conducting in-depth studies on the above-mentioned approaches. Moreover, class imbalance occurs when the minority classes contain significantly fewer EEG samples than the majority classes. When a class imbalance exists in the training data, the learned classifier overclassifies the majority classes owing to their increased prior probability. Thus, the samples belonging to minority classes are more prone to misclassification than those belonging to the majority classes. The class imbalance problem is common in several EEG applications, such as EEG seizure detection tasks. The duration of seizure events is typically much shorter than that of non-seizure periods in long-term continuous EEG data. Thus, the classifiers will be biased toward non-seizure EEG signals if the class imbalance problem is not considered. Methods for addressing class imbalance problem based on deep learning can be divided into two main categories \cite{buda2018systematic}. The first category is data level methods that operate on training set and change its class distribution \cite{buda2018systematic}. They aim to alter dataset in order to make standard training algorithms work. For example, random minority oversampling simply replicates randomly selected samples from minority classes. However, as opposed to oversampling, undersampling removes randomly from majority classes until all classes have the same number of examples. The other category covers classifier (algorithmic) level methods. These methods keep the training dataset unchanged and adjust training or inference algorithms \cite{buda2018systematic}. For example, threshold moving adjusts the decision threshold of a classifier. It is applied in the test phase and involves changing the output class probabilities. 

There are also some new problems to be solved for open-world EEG decoding. Hybrid problems are a combination of existing problems. To solve multiple issues simultaneously, several new problems will arise in open-world EEG decoding, such as few-shot transfer learning, transfer learning in the presence of noisy labels, multi-task and multi-modal transfer learning, adversarial transfer learning, adversarial training with noisy labels, self-supervised transfer learning, and multi-task zero-shot learning.

Most DL-based classification methods in EEG analysis are offline learning algorithms. However, the classifier must be retrained using all training data together with newly-arriving EEG samples, making these methods inefficient and unscalable for real-time EEG data stream analysis. Online learning can train a deep prediction model incrementally from a stream of EEG samples without requiring re-analysis of previous data, and hence it has high efficiency, strong adaptability, and excellent scalability to dynamical environments. Therefore, it is necessary to apply online learning to open-world EEG decoding, which can learn new knowledge from incoming EEG samples incrementally.

DL methods have been widely used in EEG signal analysis. The network architecture design has a significant impact on the final EEG decoding performance. Various network architectures have been designed to achieve good performance. However, network architecture design relies heavily on prior knowledge and experience. Therefore, neural architecture search \cite{elsken2019neural} aims to design the architecture of a network automatically to reduce human intervention as much as possible. Neural architecture search methods can be categorized according to three dimensions \cite{elsken2019neural}: search space, search strategy, and performance estimation strategy. The search space defines which architectures can be represented in principle. The search strategy details how to explore the search space, which is often exponentially large or even unbounded. The objective of neural architecture search is to find architectures that achieve high performance on unseen data. Performance estimation refers to the process of estimating the performance, the simplest way is to perform a standard training and validation of the architecture on data. For example, Li \emph{et al.}  proposed a novel neural architecture search framework based on reinforcement learning for EEG-based emotion recognition \cite{li2021eeg}, which can automatically design network architectures. 

The good performance of deep models for EEG decoding is at the cost of huge memory consumption and high computational complexity. There are growing interests in deploying deep models on edge devices (\emph{e.g.}, wearable device, mobile phone, medical equipment, etc.) that have a stringent budget on the resource and energy, and expect real-time processing. As a result, reducing the cost of memory and computational complexity in deep models, that is, model compression \cite{deng2020model} of deep models without significantly decreasing the model performance for EEG analysis becomes an urgent and promising topic. For example, Wang \emph{et al.} adopted the knowledge distillation to extract the distribution of training data from the complex network (teacher network) to a simple network (student network) for EEG emotion recognition \cite{wang2021fldnet}.

Current artificial intelligence for EEG decoding mainly focuses on bridging the performance gap between machines and human beings. However, general artificial intelligence replaces task-specific models with general artificial intelligence algorithmic systems, which can achieve a higher level of automation in solving diverse tasks. Automatic machine learning is a general artificial intelligence algorithm approach that can be applied to a wide range of tasks, including vastly different ones. The hyper-parameter settings of DL models have a significant impact on the final EEG decoding performance in the open world. Manual testing is a traditional and prevalent approach for tuning hyper-parameters, and it requires a deep understanding of the DL algorithms and their hyper-parameter value settings. However, manual tuning is ineffective for several problems in open-world EEG decoding. This has inspired studies on the automatic optimization of hyper-parameters. Hyper-parameter optimization aims to automate the hyper-parameter tuning process \cite{yu2020hyper}, which makes it possible for users to apply DL models to open-world EEG decoding problems effectively. In addition, when facing complex decision-making tasks in open-world EEG decoding, it is necessary to design automatic machine learning methods to solve complex tasks adaptively. Reinforcement learning is a branch of machine learning in which an agent can learn from interacting with an environment. Reinforcement learning does not require extensive engineering and heuristic design. In addition, reinforcement learning updates the parameters through trial and error, does not require the expected reward to be differentiable, and can deal with the search problem in a discrete space directly. Thus, deep reinforcement learning can combine the advantages of DL and reinforcement learning for open-world EEG decoding and hence can enable the agent to solve complex decision-making tasks.

Most deep models for EEG decoding are over-parameterized black-box models, and they can obtain high classification accuracy without interpretable knowledge representations. Therefore, it is often difficult to understand the prediction logic of deep models hidden inside the network. Thus, it is important to exploit the interpretability of deep models \cite{zhang2018visual} for EEG decoding in both theory and practice.

Most current EEG decoding methods are not comparable with human intelligence, especially in changing, dynamic, and complex open-world environments. Strong artificial intelligence aims to design algorithms that are comparable or superior to human intelligence. Therefore, it is important to develop strong artificial intelligence methods to solve the challenging problems associated with open-world EEG decoding. For example, deep models usually require a large amount of training data to achieve good performance. However, their ability to quickly learn new concepts is relatively limited. Meta-learning is known as ``learning to learn models" \cite{hospedales2021meta}. It treats tasks as training examples and aims to train a model to adapt to all such training tasks. Thus, meta-learning can improve the ability of model generalization for open-world EEG decoding, and it can potentially design general methods applicable to both in-distribution and out-of-distribution tasks.

\section{Acknowledgments}
Xun Chen, Chang Li, and Aiping Liu  were partially supported by the National Natural Science Foundation of China (grants 61922075, 41901350, and 61701158), University of Science and Technology of China
research funds from the Double First-Class Initiative (grant YD2100002004), and Anhui Huami Information Technology Co., Ltd. Martin J. McKeown was supported by John Nichol Chair in Parkinson's Research. All future correspondence should be sent to the corresponding author, Chang Li (changli@hfut.edu.cn).

\section{Authors}
\begin{IEEEbiographynophoto}{Xun Chen}
(xunchen@ustc.edu.cn) received his Ph.D. degree in biomedical engineering from the University of British Columbia. He is a distinguished professor with the Department of Neurosurgery, The First Affiliated
Hospital of USTC, Division of Life Sciences and Medicine, University of Science and Technology of China, Hefei, Anhui, 230001, China. In addition, he heads the Department of Electrical Engineering and Information Science and directs the Institute of Advanced Technology-Huami Joint Laboratory for Brain-Machine Intelligence at the University of Science and Technology of China. His research interests include signal processing and machine learning in biomedical applications. He is a Senior Member of IEEE.
\end{IEEEbiographynophoto}

\begin{IEEEbiographynophoto}{Chang Li}
(changli@hfut.edu.cn) received the B.S. degree in information and computing science from Wuhan Institute of Technology in 2012, and the Ph.D. degree in electronic information and communications from Huazhong University of Science and Technology in 2018. He is currently a Lecturer with the Department of Biomedical Engineering, Hefei University of Technology, Hefei, China. His current research interests include the areas of biomedical signal processing, hyperspectral image analysis and information fusion.
\end{IEEEbiographynophoto}

\begin{IEEEbiographynophoto}{Aiping Liu}
(aipingl@ustc.edu.cn) received her Ph.D. degree in electrical and computer engineering from the University of British Columbia. She is an associate professor in the School of Information Science and Technology, University of Science and Technology of China. Her research interests include biomedical signal processing, neuroimaging analysis, and noninvasive brain stimulation. She is a Member of IEEE.
\end{IEEEbiographynophoto}

\begin{IEEEbiographynophoto}{Martin J. McKeown}
(martin.mckeown@ubc.ca)	received his M.D. degree from the University of Toronto. He completed a three-year research fellowship in the Computational Neurobiology Laboratory at the Salk Institute for Biological Studies in San Diego, California. He was an assistant professor of medicine at Duke University from 1998 to 2003, and he is currently the John Nichol Chair in Parkinson's Research, a professor of medicine, and the director of the Pacific Parkinson's Research Center, Vancouver, British Columbia, V6E 2M6, Canada. His research interests include examining novel treatments for Parkinson's disease.
\end{IEEEbiographynophoto}

\begin{IEEEbiographynophoto}{Ruobing Qian}
(qianruobing@fsyy.ustc.edu.cn)	is the Chief Physician with the Department of Neurosurgery at The First Affiliated Hospital of University of Science and Technology of China (Anhui Provincial Hospital), the Executive Director of the Epilepsy Center, the President of the Anhui Association Against Epilepsy (AAAE). His current research interests include multimodal neuroimaging, neuroelectrophysiology, surgical treatment of refractory epilepsy, and cognitive science. He has published over 50 scientific articles in prestigious journals and conferences.
\end{IEEEbiographynophoto}

\begin{IEEEbiographynophoto}{Z. Jane Wang}
(zjanew@ece.ubc.ca)	received her Ph.D. degree in electrical engineering from the University of Connecticut. She has been a research associate at the University of Maryland, College Park from 2002 to 2004. She is a full professor in the Department of Electrical and Computer Engineering, University of British Columbia, Vancouver, British Columbia, V6T 1Z4, Canada. Her research interests include statistical signal processing and machine learning, with applications in digital media and biomedical data analytics. She is a Fellow of IEEE and the Canadian Academy of Engineers.
\end{IEEEbiographynophoto}

\bibliographystyle{IEEEtran}
\bibliography{egbib}

\end{document}